\documentstyle[psfig,twocolumn,aps,amstex,floats,graphicx,epsfig,amsfonts,enumerate]{revtex}

\tighten
\begin{document}

\draft

\title{\bf The Structure of Operators in Effective Particle-Conserving
  Models}  

\author{Christian Knetter, Kai P. Schmidt and G\"otz S. Uhrig}

\address{Institut f\"ur Theoretische Physik, Universit\"at zu
  K\"oln, Z\"ulpicher Str. 77, D-50937 K\"oln, Germany\\[1mm]
  {\rm(\today)} }

\maketitle

\begin{abstract}
For many-particle systems defined on lattices we investigate the
global structure of effective Hamiltonians and observables obtained 
by means of a suitable basis transformation. We study
transformations which lead to effective Hamiltonians conserving the
number of excitations. The {\it same} transformation must be used to
obtain effective observables. The analysis of the structure shows that
effective operators give rise to a simple and intuitive perspective
on the initial problem. The systematic calculation of $n$-particle 
irreducible quantities becomes possible constituting a significant progress.
Details how to implement the approach perturbatively for a large
class of systems are presented.
\end{abstract}


\narrowtext

\section{Introduction}
\label{intro}
Effective models are at the very center of theoretical physics
since they allow to focus on the essential physics of a problem
without being distracted by unnecessary complexity. Hence it is
very important to dispose of {\it systematic} means to derive 
effective models. Here we will present the mathematical structure
 of a certain kind of effective models, namely effective models
where the elementary excitations above the ground state
can be viewed as particles above a complex vacuum. 
This type of view is very common in low-temperature  physics.
Many experiments can be understood on the basis of  this picture.

In this paper, we will elucidate the global structure of
the Hamiltonian and of the observables if the model is transformed
to a model which {\it conserves} the number of particles. Such a mapping
is often possible and renders the subsequent calculation of 
physical quantities much easier. The determination of the effective
Hamiltonian is facilitated by the decomposition into
$n$-particle irreducible parts. We set up such a classification
at zero temperature for {\it strong-coupling} situations, i.e.\
no weak-coupling limit is needed and no non-interacting fermions or
bosons are required. Generically, we deal with hard-core bosons.

The necessity for the decomposition into $n$-particle irreducible parts
has arisen in perturbative calculations of the effective Hamiltonians because
only the $n$-particle irreducible interactions are independent of the system 
size. The second main point of this article is the perturbative 
computation of effective Hamiltonians and observables. 
Such computations are a standard 
technique for ground state energies (0-particle terms) and
 dispersion relations (1-particle terms), see
Ref.~\onlinecite{gelfa00} and references therein. But the possibility 
to compute multi-particle contributions has only recently been realized 
\cite{uhrig98c,knett00b,trebs00} and continues to be exploited
intensively. The key ingredient is to define a similarity transformation
on the operator level (see below).

A promising alternative route, which we can only sketch in this article,
 consists in the non-perturbative, renormalizing realization of the
transformation of the initial model to the effective model
which conserved the number of particles. Examples of this
approach are realized in fermionic models \cite{heidb02a,heidb02b,white02a}.

\subsection{Starting point}
We consider models which are defined on a lattice $\Gamma$. At each site
of the lattice the system can be in a number $d$ of states spanning the local 
Hilbert space. Let us assume that $d$ is finite.
The dynamics of the system is governed
by a Hamiltonian $H$ acting in the tensor-product space of the local
Hilbert spaces. For simplicity we do not consider antisymmetric, fermionic
 situations although this is also possible.
So we are focusing on  physical systems which can be described
in terms of hard-core bosons.

The Hamiltonian $H$ is assumed to be of finite range. This means
that it is composed of {\it local} operators $h_{\nu}$ acting on a 
{\it finite} number of  sites in the vicinity of the site $\nu$. 
\begin{equation}
  \label{H_local}
  H=\sum_{\nu\in\Gamma}h_{\nu}\ .
\end{equation}
We further assume, that $H$ can be split as 
\begin{equation}
  \label{H_pert}
  H(x)=U+x V\ ,
\end{equation}
so that the spectrum of $U$ is simple (see below) and that the
system does not undergo a phase transition from $x=0$ to the
range of values we are finally interested in.
These requirements do not necessarily imply that $x$ has
to be small. But it is helpful if this is the case.

The ground state of $U$ and its lowest
lying eigen-states shall be known. 
The latter will be viewed as elementary excitations
from which the whole spectrum can be built. 
We assume that we can view the elementary excitations 
above the ground state  as (quasi-)particles above the vacuum.
For simplicity, we will drop the prefix `quasi-'; it is understood
that `particle' is a synonym for elementary excitation.

We assume that the physical picture sketched for $H(x=0)=U$ is
linked {\it continuously} to the range $0\le x \le x_c$ where
$x_c$ is the critical value at which a phase transition 
occurs. At the critical value $x_c$ the picture breaks down and 
cannot be used  beyond $x=x_c$. 
Generically, a mode of $H(x)$ will become soft at  $x_c$.
 
Furthermore, the particles for $x=0$ shall be local 
 in the sense that we can assign a site to each of them.
Let $Q$ be the operator that counts the number of particles. 

As a concrete example, the reader may think of an antiferromagnetic
Heisenberg model made up from strongly coupled (coupling $J$) 
pairs of spins (`dimers') which are weakly coupled (coupling $x J$)
among themselves, e.g.~\onlinecite{knett00b,knett00a}.
At $x=0$, the ground state is the product state with singlets
on all dimers; the elementary excitations are local triplets.
The number of these local triplets, i.e.\
the number of dimers which are not in the singlet state,
 shall be given by the operator $Q$.
    
A considerable simplification of the problem can be achieved by
mapping the initial problem $H(x)$ to an effective Hamiltonian
$H_{\rm eff}(x)$ in which the number of elementary excitations does not
change. That is the number of particles should be a
conserved quantity. Then the computation of many physical quantities
is significantly simplified.

In this article, we advocate to use a continuous unitary
transformation (CUT)~\cite{wegne94,stein97,mielk98,knett00a} 
in order to achieve a systematically controlled mapping of the kind
described above which leads to
\begin{equation}
  \label{H_Q_com}
  [H_{\rm eff},Q]=0\ ,
\end{equation}
i.e.\ $H_{\rm eff}$ conserves the number of particles. Such an approach
has three major advantages:
\begin{enumerate}
\item
Conceptual clarity\\
Using a unitary transformation guarantees that no information
of the orginial model is lost. In particular,
it is clear that the {\it same} transformation 
\cite{wegne94,kehre97,knett01b,schmi01} can be applied to obtain the effective
observables  ${\mathcal O}_{\rm eff}$ from the original observables
${\mathcal  O}$. 
\item
Technical simplicity\\
To implement the unitary transformation in a continuous fashion
only the computation of commutators is required since the mapping
is split into infinitesimal steps leading to a differential equation
\cite{wegne94}
\begin{equation}
\label{flow}
\partial_\ell H = [\eta(\ell), H(\ell)]
\end{equation}
where $l\in[0,\infty]$ is an auxiliary parameter parametrizing
the continuous transformation with starting point $\ell=0$ and
end point $\ell=\infty$.
\item
Good controllability\\
By an appropriate choice of the infinitesimal generator $\eta$
of the transformation it can be designed such that is 
preserves block-band diagonality \cite{knett00a,mielk98}.
Moreover, it is renormalizing in the sense that matrix elements
between energetically very different states are transformed
more rapidly than those between energetically adjacent states
\cite{heidb02a,heidb02b,mielk98}.
\end{enumerate}

We like to point out, however, that the general structure of operators
does not depend on the details of the method by which the effective
particle-conserving model is obtained. Also other methods than CUTs
are conceivable. 

In the present paper we will focus on perturbative realizations of
the CUTs. This approach \cite{knett00a} was the first which realized
 the computation of bound states in higher orders \cite{uhrig98c,knett00b}.
The concept of a similarity transformation
is indispensible for a conceptually clear computation
of multi-particle effects \cite{trebs00,zheng01a}.

\subsection{Setup}
In Sect.~\ref{lct} we analyse global structural aspects of
effective operators. The basic prerequisite will be Eq.~(\ref{H_Q_com}).
Furthermore, we show that the linked cluster property holds.
Therefore the effective operators which hold in
the thermodynamic limit can be computed in finite systems.

Sect.~\ref{H_trafo} is a preparatory section in which
the perturbative CUT for Hamilton operators of 
certain kind is constructed. Low-dimensional spin models on lattices
are among the models which can be treated in this way.

Sect.~\ref{O_trafo} contains a detailed
description of how the perturbative CUT can be extended to transform
general observables. Series expansions in $x$ for the
effective observables are obtained which allow to compute
the experimentally relevant spectral functions.
So the extension from Hamiltonians to observables is an important one.

The article is concluded in Sect.\ V.

\section{The structure of effective operators}
\label{lct}
In this section, we {\em assume} that we are able to 
construct a mapping such that $H_{\rm eff}$ fulfills
Eq.~(\ref{H_Q_com}). The eigen-states of the particle number operator
$Q$ serve as a basis for the Hilbert space of 
the system. If the  mapping is realized perturbatively, the matrix
elements of $H_{\rm eff}$ and ${\mathcal O}_{\rm eff}$
 are polynomials in $x$. 

\subsection{The effective Hamiltonian}
\label{H_eff_general}
\subsubsection{Global structure}
 We will show that $H_{\rm eff}$ can be written as
\begin{equation}
  \label{H_hauptzerl}
  H_{\rm eff}=H_0+H_1+H_2+H_3+\ldots\ ,
\end{equation}
where $H_n$ is an $n$-particle irreducible operator, i.e. $H_n$ measures
$n$-particle energies.
 Moreover, each thermodynamic matrix element of any of the components $H_n$ 
can be obtained on finite clusters for a given order in $x$
if the original Hamiltonian is of finite range. 
The components $H_n$ can be defined recursively in 
ascending order in $n$.  

Eq.~(\ref{H_hauptzerl}) comprises already a route to determine
the properties of $H_{\rm eff}$ in a sequence of approximate
treatments. The very first step is to know the ground state
energy which defines $H_0$. The second level is to describe
the dynamics of a single particle (elementary excitations) correctly
which is possible by knowing $H_1$. The third level is reached
if $H_2$ is included which contains the information on the 
interaction of two particles. True three-particle interactions
are contained in $H_3$ and so on. From the generic experience
in condensed matter theory, the three- and more particle terms
can very often be neglected. So the first three terms in 
Eq.~(\ref{H_hauptzerl})
provide the systematically controlled starting point of a broad class
of problems.

Let us clarify some notation. We define the
following eigen-states of the particle number operator $Q$
\begin{eqnarray}
  \nonumber
  |0\rangle && \mbox{ ground state (particle vacuum)}\\\nonumber
  |i\rangle && \mbox{ state with 1 particle on site } i\\\nonumber
  |i_1i_2\rangle && \mbox{ state with 2 particles on sites }i_1\mbox{ and
    }i_2\\
\label{states} 
   & \vdots & \ ,
\end{eqnarray}
i.e.\ $Q|0\rangle=0|0\rangle$, $Q|i\rangle=1|i\rangle$ and
$Q|ij\rangle=2|ij\rangle$ and so on. These states 
span the global Hilbert space ${\mathcal E}$ of the physical system
under study. Dealing with (hard-core) bosons $|i_1i_2\rangle$
and $|i_2i_1\rangle$ are identical states. This indistinguishability
causes a certain ambiguity. This ambiguity can be remedied for instance 
by assuming that coefficients depending on several indices $i_1i_2\ldots i_n$ are even
under permutation of any pair of these indices \cite{fnote}.
 For simplicity, the  ground state $|0\rangle$ is assumed to be unique.

Let ${\mathcal R}$ be an arbitrary operator  acting on ${\mathcal
  E}$ and conserving the number of particles $[{\mathcal R},Q]=0$.  
By ${\mathcal R}|_n$ we denote the  restricted operator
acting on ${\mathcal E}_n \subset{\mathcal E}$ spanned by all
  states with exactly $n$  particles.

Now we define the operators $H_n$
\begin{mathletters}
\label{class}
\begin{eqnarray}
  H_0&:=& E_0{\mathbf 1}\\
  H_1&:=& \sum_{i;j}t_{j,i}e_{j}^{\dagger}e_{i}^
  {\phantom{\dagger}}\\
  H_2&:=& \sum_{i_1i_2;j_1j_2}\!\!\!t_{j_1j_2;i_1i_2}
  e_{j_1}^{\dagger}e_{j_2}^{\dagger}e_{i_2}^{\phantom\dagger}
  e_{i_1}^{\phantom\dagger}\\
  &\vdots & \nonumber \\
  H_n&:=& \hspace{-3mm} \sum_{i_1\ldots i_n; j_1\ldots j_n}\hspace{-3mm}
t_{j_1\ldots j_n;i_n\ldots i_1} 
e^{\dagger}_{j_1}\ldots e^{\dagger}_{j_n}
  e^{\phantom\dagger}_{i_n}\ldots e^{\phantom\dagger}_{i_1} .
\end{eqnarray}
\end{mathletters}
where ${\mathbf 1}$ is the identity operator. Note that these operators are 
defined on the full Hilbert space ${\mathcal E}$. 
The operators $e_i^{(\dagger)}$ are {\it local} operators that annihilate 
(create) particles at site $i$. They are  bosonic operators. Their definition 
can be tailored to include a hard-core repulsion between the particles to 
account for the common situation that at maximum one of the particles
may be present at given site $i$. If the particles have additional internal
quantum numbers, i.e.\ if there can be different particles at each site,
the indices $i$ and $j$ are substituted by multi-indices ${\mathbf i}$
and ${\mathbf j}$. 

As an example let us consider that there are three kinds of particles per 
sites, but that at maximum one of these particles can occupy a given site.
Then each site corresponds to a four-level system; the particles are hard-core
bosons. Such a situation arises in antiferromagnetic
dimerized spin systems where each dimer represents a four-level system. 
The ground state is the unique singlet while
the three particles are given by the three-fold degenerate triplet states.
 In this case we have the multi-indices ${\mathbf
  i}=(i,\alpha)$, where $i$ denotes the site and $\alpha$ takes for instance
the three values of the $S^z$ component $\alpha\in\{-1,0,1\}$.  In the local 
basis $\{|i,s\rangle,|i,-1\rangle,|i,0\rangle,|i,1\rangle\}$, where $s$ 
denotes the singlet, the local creation operators $e_{i,\alpha}^{\dagger}$ 
are the $4\times 4$-matrices 
\begin{mathletters}
\label{e_def}
\begin{eqnarray}
  e_{i,-1}^{\dagger} &=&\left(\begin{array}{cccc}
      0&0&0&0\\1&0&0&0\\0&0&0&0\\0&0&0&0
    \end{array}\right),\\
  e_{i,0}^{\dagger} &=& \left(\begin{array}{cccc}
      0&0&0&0\\0&0&0&0\\1&0&0&0\\0&0&0&0
    \end{array}\right),\\
    e_{i,1}^{\dagger} &=& \left(\begin{array}{cccc}
      0&0&0&0\\0&0&0&0\\0&0&0&0\\1&0&0&0
    \end{array}\right)\ .
  \end{eqnarray}
\end{mathletters}
It is understood that the action at all other sites but $i$ is the
identity so that the operators in (\ref{e_def}) are defined on the
whole Hilbert space.
The annihilation operators $e_{i,\alpha}$ are given by the hermitian
conjugate matrices. All possible commutators can easily be computed 
within the matrix representation. Finite matrix elements in the
lower right $3\times3$ block can be viewed as combined annihilation \&
creation processes: The matrix  $M_{\alpha,\beta}$ with all elements
zero except the one at $(\alpha,\beta)$ corresponds to the process
$e_{i,\alpha}^{\dagger} e_{i,\beta}^{\phantom\dagger}$. A finite
matrix element in the upper left $1\times1$ block, i.e.\ the
singlet-singlet channel, can be expressed in normal-ordered fashion as 
${\bf 1}_4 - \sum_\alpha e_{i,\alpha}^{\dagger} 
e_{i,\alpha}^{\phantom\dagger}$. In this way the operators
(\ref{e_def}) and their hermitian
conjugate define a complete algebra which in turn enables us to
classify contributions of the Hamiltonian according to the number
of particles affected as done in Eqs.~(\ref{H_hauptzerl}) and (\ref{class}).

The decomposition (\ref{H_hauptzerl}) is physically very intuitive.
Yet the next important question is whether and how the operators $H_n$ are
unambiguously defined. This issue is addressed by noting that
$H_n|_m$ vanishes for $m<n$. This follows directly from the normal-ordering
of the creation and annihilation operators in Eq.~(\ref{class}).
Then we can proceed iteratively by requiring that 
$H_{\rm eff}$ applied to $n$ particles corresponds to 
$H_0+H_1+\ldots +H_n$ ($n$ arbitrary but fixed). Solving for 
$H_n$ yields the recursions
\begin{mathletters}
\label{rekurs}
\begin{eqnarray}
  H_0|_0&:=& H_{\rm eff}|_0\\
  H_1|_1&:=& H_{\rm eff}|_1-H_0|_1\\
  H_2|_2&:=& H_{\rm eff}|_2-H_0|_2-H_1|_2\\
  &\vdots&\nonumber \\
  H_n|_n&:=& H_{\rm eff}|_n-\sum_{i=0}^{n-1}H_i|_n\ .
\label{rekurs-fin}
\end{eqnarray}
\end{mathletters}
Assuming that $H_{\rm eff}$ is calculated beforehand one starts
by evaluating $E_0$ by means of the first definition. The result
entirely defines $H_0$. The restriction $H_0|_1$ is then used in the
second equation to extract the $t_{j;i}$ of $H_1$ and so on.  
Generally, $H_n$ is defined on the full many-particle
Hilbert space, not only for $n$ particles. But it is sufficient to know
the action of $H_n$ on the subspace of $n$ particles to determine all its
matrix elements in (\ref{class}). It is the essential merit of
the notation in second quantization (\ref{class}) that it provides
the natural generalization of the action of a part of the Hamiltonian 
on a {\it finite} number of particles to an {\it arbitrary}
 number of particles.
Since Eq.~(\ref{rekurs-fin}) holds for any number of particles and since
$H_n|_m$ vanishes for $m<n$ we obtain
Eq.~(\ref{H_hauptzerl}), neglecting the precise definition of 
convergence which is beyond the scope of the present paper.

In conventional many-body language, $H_n$ stands for the $n$-particle
irreducible interaction. The subtractions in Eq.~(\ref{rekurs}) ensure
that $H_n$ contains no reducible contributions, i.e.\ contributions
which really act only on a lower number of particles. 
It should be emphasized that the formalism above does not require that a
simple {\em free} fermionic or bosonic limit exists. It is possible to
start from any type of elementary particles counted by some
operator $Q$.

Moreover, the formalism presented in this section does not
depend on how $H_{\rm eff}$ is obtained. It does not matter
whether a perturbative, a renormalizing procedure or a
rigorously exact method was used to obtain $H_{\rm eff}$.

\subsubsection{Cluster additivity}
\label{clustadditiv}
Here we focus on formal aspects of a perturbative
approach generalizing results obtained previously for
0-particle properties \cite{gelfa90} and for  1-particle properties
 \cite{gelfa96}. The feature that the Hamiltonian is of finite range
on the lattice is exploited. Then the  Eqs.~(\ref{rekurs}) can be
evaluated on finite subsystems (clusters, see below). Still,
the thermodynamically relevant matrix elements of the operators $H_n$ are 
obtained as we show in the following paragraphs.

To proceed further definitions are needed.  A {\it cluster} $C$ of the 
thermodynamic system is a {\it finite} subset of sites  of the system
and their linking bonds. By  ${\mathcal R}^C$ we denote an operator which
acts only on the Hilbert space ${\mathcal E}^C$ of $C$. If $\overline{C}$
denotes the sites of the total system which are not included in $C$, the
restricted operator ${\mathcal R}^C$ is lifted naturally to an
operator ${\mathcal R}$ in the total Hilbert space 
${\mathcal E}={\mathcal E}^C\otimes {\mathcal E}^{\overline{C}}$ by
\begin{equation}
\label{prodstruc}
{\mathcal R} := {\mathcal R}^C \otimes {\bf 1}^{\overline{C}}\ .
\end{equation}
Note that it is not possible to define a restricted operator 
${\mathcal R}^C$ from an arbitrary operator ${\mathcal R}$ acting on 
${\mathcal E}$ since ${\mathcal R}$ will not have the 
product structure (\ref{prodstruc}) in general.

Two clusters $A$ and $B$ are said to form a {\it disconnected cluster} 
$C=A\cup B$ iff they do not have any site in common $A\cap B = 0$
{\it and} there is no bond linking sites from $A$ with sites from $B$. 
Otherwise the clusters $A$ and  $B$ are said to constitute together a 
{\it linked cluster} $C=A\cup B$. 
Given a disconnected cluster $C=A\cup B$ an operator ${\mathcal R}^C$ is called
{\it cluster additive} iff it can be decomposed as
\begin{equation}
  \label{clustadd}
  {\mathcal R}^C={\mathcal R}^A\otimes{\bf 1}^B  + {\bf 1}^A\otimes
  {\mathcal R}^B \ .
\end{equation}

With these definitions we show that $H_{\rm eff}$ and $H_n$
are cluster additive. But $H_{\rm eff}|_n$ is not! This will turn out
to be another important reason to introduce the $H_n$.

The cluster additivity of $H_{\rm eff}^C$ is obvious since $A$ and 
$B$ are assumed to be disconnected. So they can be viewed as physically
independent systems. Hence
\begin{equation}
  \label{sep_1}
  H_{\rm eff}^C = H_{\rm eff}^A\otimes{\bf 1}^B + {\bf 1}^A\otimes
  H_{\rm eff}^B\ .
\end{equation}
Similarly, we deduce from (\ref{rekurs}) the operators
$H^A_n$ and $H^B_n$ which act on ${\mathcal E}^A$ and  ${\mathcal E}^B$,
respectively. Then it is straightforward to verify that the operators
\begin{equation}
H^C_n  = H_n^A\otimes{\bf 1}^B + {\bf 1}^A\otimes  H_n^B
\end{equation}
fulfil the recursion (\ref{rekurs}) for the operators defined for
the cluster $C$. Hence the operators $H_{\rm eff}$ and $H_n$ are
indeed cluster additive.

It is instructive to see that $H_{\rm eff}|_n$ is {\it not}
cluster additive, contrary to what one might have thought.
Let us consider the tentative identity
\begin{equation}
\label{wrong}
H_{\rm eff}^C|_n = H_{\rm eff}^A|_n\otimes{\bf 1}^B + {\bf 1}^A\otimes
 H_{\rm eff}^B|_n\ .
\end{equation}
This equation cannot be true since on the left hand side the number of
particles is fixed to $n$ while on the right hand side the number
of particles to which the identities ${\bf 1}^A$ and ${\bf 1}^B$
are applied is not fixed. So no cluster additivity is given for the
$H_{\rm eff}|_n$.

The fact that cluster additivity holds only for particular quantities
 was noted previously for $n=1$ \cite{gelfa96}.
For $n=2$, the subtraction procedure was first applied in the calculations
in Ref.~\onlinecite{uhrig98c} (though not given in detail). In 
Refs.~\onlinecite{knett00b,trebs00,zheng01a,zheng01b} the subtractions
necessary to obtain the irreducible 2-particle interaction were given in 
more detail. The general formalism presented in this article shows
on the {\it operator} level why such subtractions are necessary and where they 
come from. Thereby, it is possible to extend the treatment to 
 the general $n$-particle irreducible interaction. 

The notation in terms of second quantization (\ref{class}) renders 
the cluster additivity almost trivial. This is so since the creation and
annihilation operators are defined locally for a certain site. It is understood
that the other sites are not affected. Hence the same symbol $e^\dagger_i$
can be  used independent of the cluster in which the site $i$ is embedded.
In particular, one identifies automatically $e^{\dagger,C}_i$ with 
$e^{\dagger,A}_i\otimes {\bf 1}^B$ if $i\in A$ and with  
${\bf 1}^A\otimes e^{\dagger,B}_i$ if $i\in B$.
Hence cluster additivity is reduced to trivial statements of the kind that
\begin{mathletters}
\begin{eqnarray}
H_1^A &=& \sum_{i,j\in A}  t_{j;i}e_{j}^{\dagger}e_{i}^{\phantom\dagger}\\
H_1^B &=& \sum_{i,j\in B}  t_{j;i}e_{j}^{\dagger}e_{i}^{\phantom\dagger}
\end{eqnarray}
\end{mathletters}
implies
\begin{mathletters}
\begin{eqnarray}
H_1^C &=& \sum_{i,j\in C}  t_{j;i}e_{j}^{\dagger}e_{i}^{\phantom\dagger}\\
&=& \sum_{i,j\in A}  t_{j;i}e_{j}^{\dagger}e_{i}^{\phantom\dagger} +
 \sum_{i,j\in B}  t_{j;i}e_{j}^{\dagger}e_{i}^{\phantom\dagger}\\
&=& H_1^A\otimes {\bf 1}^B + {\bf 1}^A\otimes H_1^B\ .
\end{eqnarray}
\end{mathletters}
In this sense, the notation in second quantization is the most natural
way to think of cluster additivity.

Following Gelfand and co-workers \cite{gelfa00,gelfa90,gelfa96}
 we conclude that the cluster additive quantities possess a cluster expansion.
 Hence all the irreducible matrix elements
$t_{j;i}$ possess a cluster expansion and can be computed on finite clusters.

\subsubsection{Computational aspects}
Since  $H_{\rm eff}$  conserves the number of particles, 
i.e.\ Eq.~(\ref{H_Q_com}), its action
is to shift existing particles. Let us denote
 the relevant matrix elements for a linked 
cluster $A$ by
\begin{mathletters}
\label{a_def}
\begin{eqnarray}
  E_0^A &:=&\langle 0 |H_{\rm eff}^A|0\rangle\\
  a^A_{j;i} &:=& \langle j|H_{\rm eff}^A|i\rangle\\
  a^A_{j_1j_2;i_1i_2} &:=& \langle j_1j_2|H_{\rm eff}^A|i_1i_2\rangle\\
  &\vdots& \nonumber\ ,
\end{eqnarray}
\end{mathletters}
where the indices $i,j,\ldots$ may be multi-indices from now on.
Put differently, $E_0^A$ is the matrix element of $H_{\rm eff}^A|_0$,
the $a^A_{j;i}$ are the matrix elements of $H_{\rm eff}^A|_1$,
 the $a^A_{j_1j_2;i_1i_2}$ 
those of $H_{\rm eff}^A|_2$ and so on. 
The number $E_0^A$ is the ground state energy of cluster $A$. 
The recursive definitions (\ref{rekurs}) imply
\begin{mathletters}
\label{t_def}
\begin{eqnarray}
 && t_{j;i}^A = a_{j;i}^A-E_0^A\delta_{ji}\\
 && t^A_{j_1j_2;i_1i_2} =  a_{j_1j_2;i_1i_2}^A-
E_0^A\delta_{j_1i_1}\delta_{j_2i_2}-E_0^A\delta_{j_1i_2}\delta_{j_2i_1}
\nonumber\\
  && - t^A_{j_2;i_2}\delta_{j_1i_1}- t^A_{j_1;i_2}\delta_{j_2i_1}- 
t^A_{j_2;i_1}\delta_{j_1i_2}- t^A_{j_1;i_1}\delta_{j_2i_2}\\
 && t^A_{j_1j_2j_3;i_1i_2i_3} =  a^A_{j_1j_2j_3;i_1i_2i_3} - A_0 -A_1-A_2\\
  &&\vdots \ , \nonumber
\end{eqnarray}
\end{mathletters}
where $A_0$ comprises six terms resulting from $H_0$, $A_1$ comprises 
18 terms resulting from $H_1$
and $A_2$ comprises 36 terms resulting from $H_2$. The explicit
 formulae are given in Appendix \ref{3part}.
The recipe in deriving the above equations is straightforward. 
For a given $n$-particle process $\{i_m\} \to  \{j_m\}$ ($m\in\{1,\ldots,n\}$)
 one has to subtract all possible processes 
which move less than $n$ particles. Since the $m$-particle processes
with $m<n$ have been computed before the procedure is recursive. 
 Note that all coefficients must be computed for the {\it same} cluster. 

The cluster additivity or, equivalently, the existence of a cluster
expansion can be exploited to compute the irreducible matrix elements
on finite clusters given that the Hamiltonian is of finite range.
There are two strategies to do so.

The first strategy is to choose a cluster large enough to perform
the intended computation without finite-size effects. This strategy
works particularly well if the dimensionality of the problem is low.
Let us assume
for simplicity that the Hamiltonian links only nearest-neighbour sites.
Aiming at a given matrix element, for instance $t^A_{j_1j_2;i_1i_2}$,
which shall be computed in a given order $k$, the large enough
cluster $C_l$ contains all possible subcluster $C_s$ with two properties:
(i) they  have $k$ or less bonds, (ii) they link the concerned sites
$j_1,j_2,i_1,i_2$ among themselves \cite{gnote}. Clearly, $C_l$ depends on the order
$k$. But it depends also on the sites $j_1,j_2,i_1,i_2$ under  study
so that the notation $C_l^{(k)}(\{j_1,j_2,i_1,i_2\})$ is appropriate.
Note, that the order of the sites does not matter.

If some sites are omitted the constraints for the subclusters $C_s$
are diminished since less sites must be linked. This implies
in particular
$C_l^{(k)}(\{j_1,j_2,i_1,i_2\}) \subset C_l^{(k)}(\{j_1,i_1\})$.
Hence there can be  a cluster $A$ which contains
$C_l^{(k)}(\{j_1,j_2,i_1,i_2\})$ but does {\it not} contain 
$C_l^{(k)}(\{j_1,i_1\})$ so that
the hopping matrix element $t^A_{j_1;i_i}$ is {\it not} the thermodynamic
one, but the interaction $t^A_{j_1j_2;i_1i_2}$ is without
finite-size correction. So intermediate steps in the calculations
(\ref{t_def}) can display finite-size effects although the
final result does not. In Refs.~\cite{uhrig98c,knett00b,knett00a,knett01b}
we followed this strategy.

The second strategy is to compute for a given order $k$
the {\it net} contributions of all clusters $C$ with $m\le k$ bonds
which link the sites under study. The advantage of this approach
is that only smaller clusters need to be treated ($\le k$ bonds).
The price to pay is an overhead in determining the {\it net}
contribution. This requires to deduct from the total contribution
of $C$ the contributions of all subcluster of $C$ with less bonds
which link the points under study. This must be done
in order to avoid double counting.
More details on this strategy can be found in Ref.~\onlinecite{gelfa00}.

For Hamiltonians with relatively simple topology, the second strategy
is more powerful. For more complicated Hamiltonians, however, the task
to implement the overhead without flaw can quickly become impracticable
while the first strategy can still be used, at least up to a certain order
of the perturbation.

\subsection{Effective observables}
An effective Hamiltonian conserving the number of particles
is useful to determine characteristic energies of the 
considered systems. But it is not sufficient to determine
physical quantities which require more knowledge than
the eigen-energies of the system. In particular, we aim at
determining dynamic correlations such as 
$\langle {\mathcal O}(t) {\mathcal O}(0)\rangle$. Then the mapping 
of the original Hamiltonian $H$ to the effective Hamiltonian $H_{\rm eff}$
must be extended to a mapping of the original observable
 ${\mathcal O}$ to the  effective observables ${\mathcal O}_{\rm eff}$.
Here we will assume that this has been achieved by an appropriate
unitary transformation, for instance in a continuous fashion
as described in the Introduction.

\subsubsection{Global structure}
The structure of the observables can be described best 
by using the notation of second quantization. Thereby it
can be denoted clearly how many particles are involved.
The most important difference compared to the Hamiltonian
is that there is no particle conservation. Generically an observable
creates and annihilates excitations, i.e.\ particles. Hence we define
the operators  
\begin{eqnarray}
  \label{O_dn}
  &&{\mathcal O}_{d,n}:= \\ \nonumber
&&\sum_{i_1\cdots i_n;j_1\cdots j_{n+d}} 
w_{j_1\cdots j_{n+d};i_1\cdots i_n} e^{\dagger}_{j_{1}}\cdots
  e^{\dagger}_{j_{n+d}} e^{\phantom\dagger}_{i_n}\cdots
  e^{\phantom\dagger}_{i_1} .
\end{eqnarray}
The local operators $e_i$ have been described after Eq.~(\ref{class}). Again 
they shall appear normal-ordered, i.e.\ all creation operators are sorted  
to the left 
of the annihilation operators. The first index $d$ indicates how many 
particles are created ($d\ge 0$) or annihilated ($d<0$) by application of
 ${\mathcal O}_{d,n}$. The second index $n\ge0$ denotes how many particles
have to be present before the operator ${\mathcal O}_{d,n}$ becomes active.
The result of ${\mathcal O}_{d,n}$ acting on a state
with less than $n$ particles is zero.

In analogy to Eq.~(\ref{H_hauptzerl}) the effective observables can be 
decomposed into partial observables like
\begin{equation}
  \label{O_hauptzerl}
  {\mathcal O}_{\rm eff}=\sum_{n=0}^{\infty}\ \sum_{d\ge -n} {\mathcal
  O}_{d,n}\ .
\end{equation}
The additional feature in comparison to Eq.~(\ref{H_hauptzerl}) is the
sum over $d$. Tab.~\ref{O_zersplittert} sketches the structure of
the terms appearing in the partial observables ${\mathcal O}_{d,n}$
\begin{table}[ht]
\begin{tabular}{|c||c|c|c|c|}
\hline
$d\downarrow$ / $n\rightarrow$ & 0 & 1 & 2 & 3 \\
\hline\hline
$\cdots$ & $\cdots$ & $\cdots$ & $\cdots$ & $\cdots$ \\
\hline
-3 &$0$&$0$&$0$&$eee$ \\
\hline
-2 &$0$&$0$&$ee$&$e^{\dagger}eee$ \\
\hline
-1 &$0$&$e$&$e^{\dagger}ee$&$e^{\dagger}e^{\dagger}eee$ \\
\hline
0 &$1$&$e^{\dagger}e$&$e^{\dagger}e^{\dagger}ee$&$e^{\dagger}e^{\dagger}e^{\dagger}eee$ \\
\hline
1 &$e^{\dagger}$&$e^{\dagger}e^{\dagger}e$&$e^{\dagger}e^{\dagger}e^{\dagger}ee$&$e^{\dagger}e^{\dagger}e^{\dagger}e^{\dagger}eee$ \\
\hline
2 &$e^{\dagger}e^{\dagger}$&$e^{\dagger}e^{\dagger}e^{\dagger}e$&$e^{\dagger}e^{\dagger}e^{\dagger}e^{\dagger}ee$&$e^{\dagger}e^{\dagger}e^{\dagger}e^{\dagger}e^{\dagger}eee$ \\
\hline
$\cdots$ & $\cdots$ & $\cdots$ & $\cdots$ & $\cdots$ \\
\hline
\end{tabular}
\caption{\label{O_zersplittert} List of terms appearing in the
  partial observables ${\mathcal O}_{d,n}$ which form together the
  effective observable ${\mathcal O}_{\rm eff}$ according to
  Eq.~(\ref{O_hauptzerl}). No prefactors or indices are given for clarity.}  
\end{table}

Let us assume that we computed ${\mathcal O}_{\rm eff}$ by some technique,
for instance by a CUT. Then the partial observables can be determined
recursively by
\begin{mathletters}
\label{O_rekurs}
\begin{eqnarray}
\label{O_rekursa}
  {\mathcal O}_{d,0}|_{0\rightarrow 0+d}&:=& {\mathcal O}_{\rm
    eff}|_{0\rightarrow 0+d}\\
  {\mathcal O}_{d,1}|_{1\rightarrow 1+d} &:=& {\mathcal O}_{\rm
    eff}|_{1\rightarrow 1+d}- {\mathcal O}_{d,0}|_{1\rightarrow
    1+d}\\ \nonumber 
  {\mathcal O}_{d,2}|_{2\rightarrow 2+d} &:=& {\mathcal O}_{\rm
    eff}|_{2\rightarrow 2+d}- {\mathcal O}_{d,0}|_{2\rightarrow
    2+d}-{\mathcal O}_{d,1}|_{2\rightarrow 2+d}\\\nonumber
  &\vdots& \\
  {\mathcal O}_{d,n}|_{n\rightarrow n+d} &:=& {\mathcal O}_{\rm
    eff}|_{n\rightarrow n+d} - \sum_{i=0}^{n-1}{\mathcal
  O}_{d,i}|_{n\rightarrow n+d}\ .
\end{eqnarray}
\end{mathletters}
Here $|_{n\rightarrow n+d}$ denotes the restriction of an operator to
act on the $n$-particle subspace ${\mathcal E}_n$ (domain) and to yield
states in the $(n$+$d)$-particle subspace ${\mathcal E}_{n+d}$ (co-domain). 
The recursion is set-up in analogy to (\ref{rekurs}). It is again used
that an operator ${\mathcal O}_{d,n}$ effectively vanishes if it is applied
to {\it less} than $n$ particles. Barring possible problems to define
convergence, the validity of the recursion (\ref{O_rekurs}) for
all $d$ and $n$ implies the decomposition (\ref{O_hauptzerl}).

As for the Hamiltonian the partial observables ${\mathcal O}_{d,n}$ 
can be viewed as the $n$-particle irreducible part of the particular
observable. The notation in second quantization elegantly resolves the
question how the observables act on clusters as was explained in the
section \ref{clustadditiv}. Hence the definition (\ref{O_dn}) ensures
cluster additivity and there exist cluster expansions for
the partial observables. So they can be computed on finite clusters.

If dynamical correlations at zero temperature $T=0$ shall be described,
the observables are applied to the ground state $|0\rangle$ which is
the particle vacuum \cite{heidb02b}. Then only the partial observables
 ${\mathcal O}_{d,0}$ with $d\ge0$ matter. According
to (\ref{O_rekursa}) no corrections are necessary, i.e.\
the structure of the relevant part of the effective observable
is given by
\begin{equation}
  \label{O_d0_extra}
  {\mathcal O}_{\rm eff}^{T=0}={\mathcal O}_{0,0}+{\mathcal
  O}_{1,0}+{\mathcal O}_{2,0}+{\mathcal O}_{3,0}+\ldots\ .
\end{equation}
This structure has been used so far in a number of investigations
of spectral weights \cite{singh99a,schmi03a} and spectral densities
\cite{knett01b,schmi01,zheng03a}. It turned out that it is indeed
sufficient to  consider a restricted number of particles 
\cite{knett01b,schmi01,schmi03a}. But the question how many particles
are required to describe a certain physical quantity sufficiently well 
depends on the considered model, the chosen basis (What do we call a particle?)
and the quantity under study.

At finite temperatures a certain number of particles will already be
present in the system due to thermal fluctuations. Then the action of
the partial observables ${\mathcal O}_{d,n}$ with $n\ge 1$ will
come into play as well. This constitutes an interesting route
to extend the applicability of effective models, which were derived
in the first place at zero temperature, to finite temperatures.

\subsubsection{Computational aspects}
The recursive equations for matrix elements which can be derived
from (\ref{O_rekurs}) are very similar to those obtained
for the Hamiltonian (\ref{t_def}). We illustrate this for
the matrix elements of ${\mathcal O}_{1,n}$. Let the bare
matrix elements on a cluster $A$ be
\begin{mathletters}
\label{v_def}
\begin{eqnarray}
  v^A_j &:=&\langle j|{\mathcal O}_{\rm eff}^A|0\rangle\\
  v^A_{j_1j_2;i} &:=&\langle j_1j_2|{\mathcal O}_{\rm eff}^A|i\rangle\\
\nonumber & \vdots &\ .
\end{eqnarray}
\end{mathletters}
From (\ref{O_rekurs}) we obtain the irreducible elements as
\begin{mathletters}
\label{w_def}
\begin{eqnarray}
  w_j^A &= &v_j^A\\
 w^A_{j_1j_2;i} &:=&v^A_{j_1j_2;i}-w_{j_1}^A \delta_{j_2i}
-w_{j_2}^A \delta_{j_1i}\\
\nonumber & \vdots &\ .
\end{eqnarray}
\end{mathletters}
As for the irreducible interactions the strategy is straightforward. 
One has to subtract from the reducible $n$-particle  matrix elements $v^A$
the contributions which come from the $m$-particle irreducible
 matrix elements $w^A$ with $m<n$.
With this strategy also other irreducible
matrix elements can be determined in a straightforward manner.

So far our considerations were general in the sense that it did not
matter how we achieved the mapping.
Next we focus on the actual {\it perturbative}
 evaluation of the matrix elements on
finite clusters. For simplicity, we assume as before that the 
perturbative part of the Hamiltonian links only nearest-neighbour
sites. Let us consider for instance $w^A_{j_1j_2;i}$. We assume that
the observable ${\mathcal O}$ is also local, i.e.\ acts on 
a certain site only, or is a sum of such terms. If the observable is
 a sum of local terms then the transformation of each term separately
and subsequent summation yields the result. So without loss of generality
we consider ${\mathcal O}$ to affect only site $p$. Then we have to compute the
matrix elements for clusters linking the {\it four} sites $j_1,j_2,i,p$.
If ${\mathcal O}$ itself is a product of operators affecting several
sites $p_i$ then the observable ${\mathcal O}$ itself links these sites
$p_i$. Apart from this difference compared to the matrix elements of
the effective Hamiltonian,
we may copy the remaining steps from there:

There are again the two strategies. Either the calculation in order $k$
 is performed on a cluster $C_l$ large enough so that all subclusters of
$k$ bonds linking the relevant sites $j_1,j_2,i,p$ are comprised in $C_l$
\cite{knett01b,schmi01,knett02,gruni02b}.
Or one has to add the {\it net} contributions of all different
clusters with $k$ or less bonds which link the relevant sites $j_1,j_2,i,p$
\cite{zheng03a}.
In either way the results for spectral densities 
can be obtained.

\section{Transformation of the Hamiltonian}
\label{H_trafo}
So far no particular property of the transformation providing
 the effective operators $H_{\rm eff}$ and
${\mathcal O}_{\rm eff}$ was assumed. The only prerequisites were
the existence of a counting operator $Q$, which counts the number 
of elementary excitations, i.e.\ particles, and the conservation of this
number of particles by $H_{\rm eff}$: $[H_{\rm eff},Q]=0$.

Here we specify a particular transformation
leading to  $[H_{\rm eff},Q]=0$. This section is a very brief
summary of Ref.~\onlinecite{knett00a} which is necessary to 
fix the ideas and the notation for the subsequent section
dealing with the transformation yielding the effective observables.

\begin{figure}[ht]      
\begin{center}  
    \includegraphics[width=5cm]{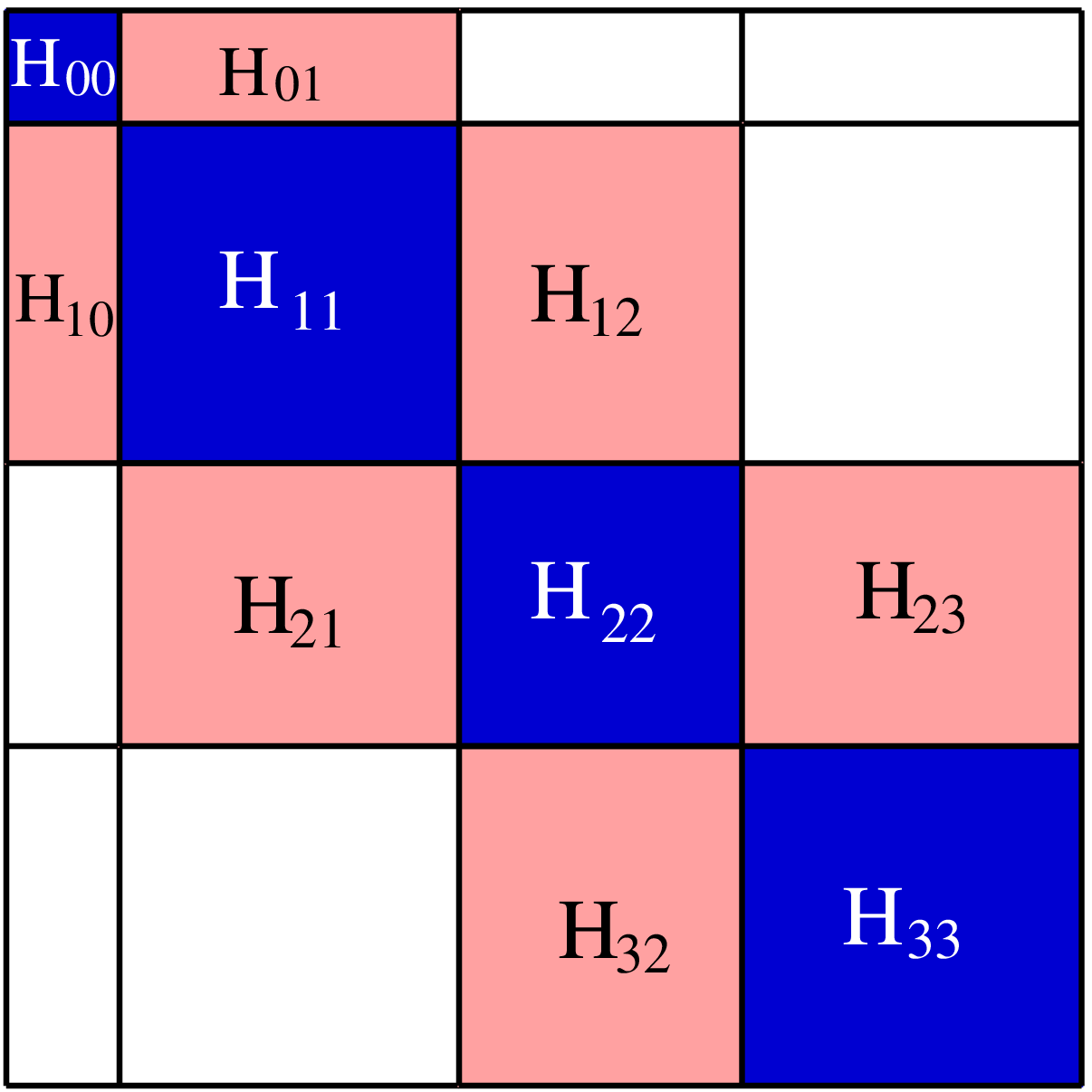}
    \caption{\label{H-Matrix} Block-band diagonal 
      Hamilton matrix for $N=1$ in the eigen-basis 
      $\{|n\rangle\}$ of the operator $Q$ which counts the
      number of particles. The unperturbed Hamiltonian
      $H(x=0)=U$ and the effective Hamiltonian $H_{\rm eff}$ have
      matrix elements in the dark areas only: $[H_{\rm eff},Q]=0$.
      For non-degenerate ground state  $H_{00}$ is 
      a $1\times 1$ matrix. The dimension of $H_{nn}$ grows roughly
      like $L^n$ with system size $L$. The perturbation $V$
      can lead to overlap matrices indicated as light boxes. The empty
      boxes contain vanishing matrix elements only.}
  \end{center}
\end{figure}

 For simplicity we restrict the
considered systems in the following way: The problem can be formulated
as perturbation problem as in Eq.~(\ref{H_pert}) with the properties 
\begin{enumerate}[(A)]
\item The unperturbed part $U$ has an
  equidistant spectrum bounded from below. The difference between two
  successive levels is the energy of a particle, i.e.\ $Q=U$.
\item There is a number ${\mathbb N} \ni N>0$ such that the perturbing
  part $V$ can be split as $V=\sum_{n=-N}^{N}T_n$ where $T_n$
  increments (or decrements, if $n<0$) the number of particles by $n$:
  $[Q,T_n]=nT_n$.
\end{enumerate}
Condition (A) allows to introduce the particularly simple and intuitive
choice  $Q=U$. Note that the restrictions of (A) are not too serious 
in practice since very often the deviations from an equidistant
spectrum can be put into the perturbation $V$.
Conditions (A) and (B) together imply that the
starting Hamiltonian $H$ has a block-band-diagonal structure
 as depicted in Fig.~\ref{H-Matrix}. The
perturbation $V$ connects states of different particle numbers only if
the difference is a finite number $\le N$. Note that very many
problems in physics display this property, for a discussion of
interacting fermions see Ref.~\onlinecite{heidb02a,heidb02b}.
So far, most applications consider $N=1$ \cite{knett00b,knett00e}  and $N=2$ 
\cite{uhrig98c,knett00a,stein97,knett01b,schmi01,mulle00a,schmi03a,knett02,gruni02b,knett01a},
but calculations for higher $N$ are also possible \cite{breni02}.

We solve the flow equation (\ref{flow}) for the Hamiltonian (\ref{H_pert})
obeying the conditions (A) and (B) perturbatively, that means up to a
certain order in the expansion parameter $x$. The  {\it ansatz} used is
  \begin{equation}
    \label{H_ansatz}
    H(x;\ell) = U +\sum_{k=1}^{\infty}x^{k} 
\sum_{|\underline{m}|=k} F(\ell;\underline{m}) 
T(\underline{m}) , 
  \end{equation}
with unknown real functions $F(\ell;\underline{m})$ for which the flow
equation (\ref{flow}) yields non-linear recursive differential
equations \cite{knett00a}. The notation comprises
\begin{mathletters}
\label{sym_def} 
\begin{eqnarray}
  {\underline m}&=&(m_1,m_2,m_3,\ldots,m_k)\quad \mbox{with}\\
  m_i &\in& \{0,\pm 1,\pm 2,\ldots ,\pm N\}\\
  |\underline{m}|&=&k\\
  T({\underline m})&=&T_{m_1}T_{m_2}T_{m_3}\cdots
  T_{m_k}\\
  M({\underline m})&=&\sum_{i=1}^km_i\ .
\end{eqnarray}
\end{mathletters}
The second sum in ansatz (\ref{H_ansatz}) runs over all indices
${\underline m}$ of length $|{\underline m}|=k$. Thereby, $H(x;\ell)$ includes all possible virtual excitation processes $T({\underline m})$ 
in a given order $x^k$  multiplied by the weight $F(\ell;{\underline m})$.   

The optimum choice for the infinitesimal
generator $\eta$ of the unitary transformation reads
\begin{equation}
  \label{eta_bfree}
  \eta(x;\ell)=\sum_{k=1}^{\infty}x^k\sum_{|{\underline m}|=k}
  \mbox{sgn}\left(M({\underline m})\right)F(\ell;{\underline
  m})T({\underline m}) .
\end{equation}
In the eigen-basis $\{|n\rangle\}$ of $Q$,
i.e.\ $Q|n\rangle=n|n\rangle$, the matrix elements of the 
generator $\eta$ read
  \begin{equation}
    \label{gen}
    \eta_{i,j}(x;\ell)={\rm sgn} (Q_i-Q_j)H_{i,j}(x;\ell)\ ,
  \end{equation}
with the convention ${\rm sgn}(0)=0$. This choice keeps the
 the flowing Hamiltonian block-band diagonal
also at intermediate values of $\ell$ \cite{knett00a,mielk98}.
For $\ell \rightarrow \infty$ the generator (\ref{gen}) eliminates all
parts of $H(x;\ell)$ changing the number of particles so that
 $[H_{\rm eff},Q]=0$ with $H_{\rm eff}:=H(\ell=\infty)$.

For the  functions $F(\ell;\underline{m})$  a set of 
coupled differential equations is determined by
inserting Eqs.~(\ref{H_ansatz}) and (\ref{eta_bfree}) in the flow
equation (\ref{flow}) and comparing coefficients. The differential
equations are recursive \cite{knett00a}. The functions $F$ of order $k+1$,
i.e.\ $F(\ell;{\underline m})$ with $|{\underline m}|=k+1$, are
determined by the functions $F$ of order $k$. 
The initial conditions are $F(0;{\underline m})=1$ for $|{\underline m}|=1$
and $F(0;{\underline m})=0$ for $|{\underline m}|>1$.
The functions are sums of monomials with structure
$(p/q)\ell^i\exp(-2\mu \ell)$, where $p,q,i,(\mu>0)$ are integers. 
This allows to implement a computer-aided iterative algorithm for the
 computation of the functions $F$ \cite{knett00a}. 

The following symmetry relations hold
\begin{mathletters}
\label{F_sym}
\begin{eqnarray}
  \label{sym1}
  F(\ell;{\underline m})&=&F(\ell;(-m_k,\ldots,-m_1))\\
  F(\ell;{\underline m})&=&F(\ell;(-m_1,\ldots,-m_k))(-1)^{|{\underline m}|+1}
\ .
\end{eqnarray}
\end{mathletters}
Relation (\ref{sym1}) reflects the hermitecity of the Hamiltonian.
The block-band diagonality  for all $\ell$ implies
\begin{equation}
\label{F_block}
F(\ell;{\underline m})=0\quad \mbox{for}\quad  |M(\underline m)|>N\ .
\end{equation}

In the limit $\ell \rightarrow \infty$  the coefficients $C(\underline{m}):=
F(\infty;\underline{m})$ are obtained. They are available in paper form
\cite{knett00a,knett00e} and electronically \cite{notiz1}.
The effective Hamiltonian is given by the general form
\begin{equation}
  \label{H_eff}
  H_{\rm eff}(x) = U +\sum_{k=1}^{\infty}x^{k} 
\sum_{|\underline{m}|=k \atop M(\underline{m})=0}  C(\underline{m}) 
T(\underline{m})\ ,
\end{equation}
where $M(\underline{m})=0$ reflects the conservation of the number of 
particles. The action of  $H_{\rm eff}$ can be viewed as a weighted sum 
of particle-number conserving virtual excitation processes each of which
is encoded in a monomial $T(\underline{m})$. 
We want to emphasize that the effective Hamiltonian $H_{\rm
  eff}$ with known coefficients $C(\underline{m})$ 
can be used straightforwardly in all perturbative problems that meet
conditions (A) and (B).

\section{Transformation of observables}
\label{O_trafo}
To calculate physical quantities which do not only depend on the
eigen-energies the relevant observables must also be known.
The conceptual simplicity of unitary transformations implies that
the observables must be
subject to the {\it same} unitary transformation as the
Hamiltonian. In this section we describe how the perturbative CUT
method can be extended to serve this purpose.

Consider the observable ${\mathcal O}$.  It is mapped according
to the flow equation 
\begin{equation}
  \label{Q_flow}
  \frac{\partial {\mathcal O}(x;\ell)}{\partial
  \ell}=[\eta(x;\ell),{\mathcal O}(x;\ell)]\ ,
\end{equation}
where the {\it same} generator $\eta(x;\ell)$, given in
Eq.\ (\ref{eta_bfree}), as in Eq.~(\ref{flow})
is to be used  to generate the transformation. 
 In analogy to Eq.~(\ref{H_ansatz}) we employ the ansatz 
\begin{equation}
  \label{O_ansatz}
  {\mathcal O}(x;\ell)=\sum_{k=0}^{\infty}x^k\sum_{i=1}^{k+1}
  \sum_{|\underline{m}|=k}G(\ell;\underline{m};i){\mathcal
  O}(\underline{m};i) ,
\end{equation}
where the $G(\ell;\underline{m};i)$ are real-valued functions for
which the flow equation (\ref{Q_flow}) yields recursive
differential equations. The operator products
${\mathcal O}(\underline{m};i)$ are given by
\begin{equation}
  \label{O_mi_def}
  {\mathcal O}(\underline{m};i):=T_{m_1}\cdots T_{m_{i-1}} {\mathcal O}
T_{m_{i}} \cdots T_{m_{k}}\ ,
\end{equation}
where we use the notation of the Eqs.~(\ref{sym_def}).
 The integer $i$ denotes the position in
${\mathcal O}({\underline m},i)$ at which the operator ${\mathcal O}$
is inserted in the sequence of the $T_m$. The starting condition
is ${\mathcal O}(x;0)= {\mathcal O}(x)$ and the final result is found at
$\ell=\infty$: ${\mathcal O}_{\rm eff}(x):={\mathcal O}(x;\infty)$.  

Inserting the ansatz (\ref{O_mi_def}) for ${\mathcal O}(x;\ell)$  and the 
generator $\eta(x;\ell)$ from (\ref{eta_bfree}) into
 the flow equation (\ref{Q_flow}) yields
\begin{eqnarray}
  \nonumber
  &&\sum_{k=0}^{\infty}x^k\sum_{|{\underline m}|=k}\sum_{i=1}^{k+1}
  \frac{\partial}{\partial \ell}G(\ell;{\underline m};i) {\mathcal
    O}({\underline m};i)=\\\nonumber
  &&\hspace*{10mm} \sum_{k_1=1}^{\infty}\sum_{k_2=0}^{\infty}x^{k_1+k_2}\!\!\!\!\!\!
  \sum_{|{\underline 
  m}'|=k_1 \atop |{\underline m}''|=k_2}\!\!\sum_{i=1}^{k_2+1}
  F(\ell;{\underline m}')G(\ell;{\underline m}'';i)\times\\\label{G_dgl_1}
  && \hspace*{20mm}\times\mbox{sgn}
  (M({\underline m}'))\left[T({\underline m}'),{\mathcal O}
  ({\underline m}'';i)\right]\ .
\end{eqnarray} 
The functions $F(\ell;{\underline m})$ are known from the
calculations described in the previous section \ref{H_trafo}
pertaining to the transformation of the Hamiltonian. 
The  sums denoted by expressions of the type $|\underline m| =k$  run over all
 multi-indices $\underline m$ of length $k$.

Comparing coefficients in Eq.~(\ref{G_dgl_1}) yields  a set of
recursive differential  equations for  the functions
$G(\ell;{\underline m},i)$.  To ease the comparison of coefficients
we split a specific $\underline m$ with $k$
fixed in two parts as defined by $i$
\begin{equation}
  \label{m_split}
  {\underline m}=({\underline m}_l,{\underline m}_r)\ ,
\end{equation}
with $|{\underline m}_l|=i-1$ and $|{\underline m}_r|=k-i+1$
such that the splitting reflects the structure of 
 ${\mathcal O}({\underline m};i)$ in (\ref{O_mi_def}).
Then the exlicit recursions can be denoted by
\begin{eqnarray}
  \nonumber
  && \frac{\partial}{\partial \ell} G(\ell;{\underline m};i)=\\\nonumber
  && \sum_{{\underline m}_l=({\underline m}_a,{\underline m}_b)
        \atop {\underline m}_a\not=0} \hspace{-2mm} \mbox{sgn}(M({\underline
  m}_a))F(\ell;{\underline m}_a) G(\ell;({\underline m}_b,{\underline
  m}_r);i-|{\underline  m}_a|)\\
  &&- \sum_{{\underline m}_r=({\underline m}_a,{\underline m}_b)
        \atop {\underline m}_b\not=0} \hspace{-2mm} \mbox{sgn}(M({\underline
        m}_b))F(\ell;{\underline m}_b)
   G(\ell;({\underline m}_l,{\underline m}_a);i)\ .\label{G_dgl_2}
\end{eqnarray}
The recursive nature of these equations becomes apparent by observing
that the summations ${\underline m}_l=({\underline m}_a,{\underline
  m}_b)$ and ${\underline m}_r=({\underline m}_a,{\underline m}_b)$
are performed over all non-trivial breakups of ${\underline
  m}_l$ and ${\underline m}_r$. For instance, the restriction ${\underline
  m}_l=(m_1,m_2,\ldots,m_{i-1})\dot=({\underline m}_a,{\underline m}_b)$
with ${\underline m}_a\not =0$ means, that one has to sum over the breakups
\begin{alignat}{3}
  \nonumber
  &{\underline m}_a=(m_1) & \mbox{ \ and \ } &{\underline
    m}_b=(m_2,\ldots,m_{i-1})\\\nonumber 
  &{\underline m}_a=(m_1,m_2) & \mbox{ \ and \ } &{\underline
    m}_b=(m_3,\ldots,m_{i-1})\\\nonumber
  &\vdots  & &\vdots \\\label{m_break_2}
  &{\underline m}_a=(m_1,m_2,\ldots,m_{i-1}) & \mbox{ \ and \ } &{\underline
    m}_b=()\ .
\end{alignat}
This implies that the $G(\ell;{\underline m};i)$ appearing on the
right side of Eq.~(\ref{G_dgl_2}) are of order  $k-1$ or less. Once they
are known the function on the left hand side of order $k$ can be computed.
By iteration, all functions can be determined.
The initial conditions follow from 
${\mathcal O}(x;\ell=0)={\mathcal O}$ and read 
\begin{mathletters}
\label{start_cond}
\begin{eqnarray}
  G(0;{\underline m};1)&= & 1\mbox{ for }|{\underline
    m}|= 0\\
  G(0;{\underline m};i)&= & 0\mbox{ for }|{\underline
    m}|> 0\ .
\end{eqnarray}
\end{mathletters}
By iteration of (\ref{G_dgl_2}), all functions can be determined.

We briefly discuss two examples to illustrate how the 
 Eqs.~(\ref{G_dgl_2}) work. Let us assume $N=2$.
 All zero order functions $G(\ell;(),1)$ are
equal to $1$. Since there is no breakup of $()$, as would be
required by the sums
on the right hand side of Eqs.~(\ref{G_dgl_2}), the right hand sides
vanish identically, whence $G(\ell;();1)=1$ for all values of $\ell$. 

The first order function $G(\ell;(1);2)$ is given by 
\begin{eqnarray}\nonumber
  \frac{\partial}{\partial
  \ell} G(\ell;(\underbrace{1}_{m_l});2)
        &=&{\rm sgn}\left[M((1))\right]F(\ell;(1))\cdot G(\ell;();1)\\ 
  &=& e^{-\ell}\cdot 1\ ,\label{example_1}
\end{eqnarray}
where $F(\ell;(1))=e^{-\ell}$ is taken from
Eq.~(15) in Ref.~\onlinecite{knett00a}. With the initial condition
$G(0;(1);2)=0$  from (\ref{start_cond}) the differential
equation (\ref{example_1}) yields
\begin{equation}
  \label{ex_1_res}
  G(\ell;(1);2)=1-e^{-l}@>>{\ell\rightarrow\infty}>1\ .
\end{equation}

As a second example we consider a second order function where we can
use the above result
\begin{mathletters}
\label{example_2}
\begin{eqnarray}
  \nonumber
  \frac{\partial}{\partial \ell}
  &&G(\ell;(\underbrace{-2,1}_{m_l});3)=\\\nonumber
  &&\hspace*{2mm}{\rm sgn}\left[M((-2,1))\right]F(\ell;(-2,1))
    G(\ell;(),1)\\
  &&\hspace*{5mm}+{\rm sgn}\left[M((-2))\right]F(\ell;(-2))
  G(\ell;(1),2)\\
  &&=-\left(e^{-3\ell}-e^{-\ell}\right)\cdot 1 - e^{-2\ell}\cdot
  \left(1-e^{-\ell}\right)\\
  &&=e^{-\ell}-e^{-2\ell}\label{ex-fin-res}
\ .
\end{eqnarray}
\end{mathletters}
Again the functions $F$ are taken from Eq.~(15) in Ref.~\onlinecite{knett00a}.
Integrating the result (\ref{ex-fin-res}) using the initial
condition (\ref{start_cond}) leads to
\begin{equation}
  \label{ex_2_res}
  G(\ell;(-2,1);3)=-e^{-\ell}+\textstyle\frac{1}{2}e^{-2\ell}
+1-\textstyle\frac{1}{2}
  @>>{l\rightarrow\infty}>\textstyle\frac{1}{2} .
\end{equation}
This kind of calculation carries forward to higher orders. The
functions $G$ -- like the functions $F$ -- are sums of simple monomials  
$(p/q)\ell^i\exp(-2\mu \ell)$, where $p,q,i,(\mu>0)$ are integers. 
 Thus the  integrations are always straightforward
\begin{mathletters}
\begin{eqnarray}
  \int_0^{\ell} d\ell' {\ell'}^i &=& {\textstyle\frac{1}{i+1}}\ell^{i+1}\\
  \int_0^{\ell} d\ell' {\ell'}^i e^{-2\mu \ell'} &=& 
{\frac{i!}{2\mu}}\left[{\textstyle\frac{1}{(2\mu)^i}}-
e^{-2\mu \ell}\sum_{j=0}^i\textstyle\frac{\ell^j}{j! (2\mu)^{i-j}}\right]
\end{eqnarray}
\end{mathletters}
and can easily be implemented in a computer-algebraic programme.
The remaining implementation follows very much the same line as
described previously for the functions $F$ \cite{knett00a}.

In analogy to Eqs.~(\ref{F_sym}) for $F$  two symmetry relations hold for $G$.
With ${\underline m}=(m_1,\ldots,m_k)$ they read
\begin{mathletters}
\label{G_sym}
\begin{eqnarray}\label{Osym1}
  G(\ell;{\underline m};i)&=&
  G(\ell;(-m_k,\ldots,-m_1);k-i+2)\\\label{Osym2}
  G(\ell;{\underline m};i)&=&
  G(\ell;(-m_1,\ldots,-m_k);i)(-1)^{|\underline m|}
\end{eqnarray} 
\end{mathletters}
as can be shown by induction.
The first symmetry (\ref{Osym1}) holds if  ${\mathcal O}$ is hermitian. 
Unfortunately, there is no equivalence to Eq.~(\ref{F_block}) so that a
possible initial block-band structure in ${\mathcal O}(x;0)$ is generically
 lost in the course of the transformation, i.e.\ for $\ell >0$.

In the limit $\ell\rightarrow\infty$ the coefficients
$\tilde{C}(\underline{m};i):=G(\infty;\underline{m};i)\in\mathbb Q$
 are obtained as rational numbers. So we retrieve finally
\begin{equation}
  \label{O_eff_1}
  {\mathcal O_{\rm eff}}(x) = \sum_{k=0}^{\infty}x^k\sum_{i=1}^{k+1}
  \sum_{|\underline{m}|=k}  \tilde{C}(\underline{m};i){\mathcal
  O}(\underline{m};i)
\end{equation} 
similar to Eq.~(\ref{H_eff}).
We will make the coefficients $\tilde{C}(\underline{m};i)$
available electronically \cite{notiz1}. Note that ${\mathcal O}_{\rm eff}$ is
{\it not} a  particle-conserving quantity as is obvious from
the fact that the sum over $|\underline{m}|$ is not restricted to
$M(\underline{m})=0$. In order to see the net effect of 
${\mathcal O_{\rm eff}}(x)$ on the number of particles explicitly
it is helpful to split the bare operator accordingly 
${\mathcal O}=\sum_{n=-N'}^{N'}T'_n$, where $T'_n$ increments (or decrements, 
if $n<0$) the number of particles by $n$: $[Q,T'_n]=nT'_n$.

The difference between the bare initial observable
${\mathcal O}$ and the representation (\ref{O_eff_1}) must be viewed
as vertex correction which comes into play since the bare initial excitations
are not the true eigen-excitations of the interacting system.
We like to stress that the formalism presented introduces the
notions of $n$-particle irreducibility, vertex correction and so on
{\it without} starting from the limit of {\it non-interacting}
 conventional particles such as bosons or fermions. 

\section{Conclusions} 
\subsection{Summary}
In this article we have presented an approach  to
calculate energies and observables for quantum multi-particle systems
defined on lattices. The article has two main parts. In the
first part (Sect.\ II), we assumed the existence of a mapping 
of the original problem to an effective one in which the number
of elementary excitations, the so-called (quasi-)particles,
is conserved. The general structure of the effective Hamiltonians
and the observables is analysed. We found that a classification
of the various contributions in terms of the number of particles
concerned is most advantageous. To this end we introduced a
notation in second quantization which does not, however, require
non-interacting fermions or bosons. Generically, hard-core bosons
are involved. 

We found the formulation in second quantization particularly intuitive.
It provides in a natural way the irreducible quantities on the 
{\it operator level} which display cluster additivity.
We like to emphasize that the definition of irreducible operators
is not a trivial task if a strong-coupling situation is considered
as was done in the present paper. No limit of non-interacting 
bosons or fermions is assumed.
Since the definition of irreducible operators is completely
general it allows to compute the $n$-particle contribution
for arbitrary $n$. For instance the formulae for the $3$-particle
interactions are given for the first time in the literature.

The {\it irreducible} interactions and vertex corrections possess
a cluster expansion so that they can be computed on finite
clusters provided that the Hamiltonian is of finite range.
This property is the basis for the real-space treatment of many spin systems.

In the second part (Sects.\ III and IV), we described 
an actual mapping which provides effective operators. The mapping
is based on continuous unitary transformations. In this
paper we constructed the mapping perturbatively (see `Outlook').
In Sect.\ III the treatment of the Hamiltonian is given. The computation 
of the effective Hamiltonian requires the solution of a set of recursive
non-linear differential equations. For the perturbative set-up
under study these equations can be solved in full generality, i.e.\
no particular details of the model must be known.

In  Sect.\ IV we have given the calculational steps to compute 
effective observables. Again recursive differential equations
have to be solved. But they are linear since the 
transformation of the Hamiltonian is known. For the perturbative set-up
under study also the equations for the observables
can be solved in full generality, i.e.\
no particular details of the model must be known.

The above approach has been used to compute spectral
functions, i.e.\ dynamical correlations, in a number of models 
\cite{uhrig98c,knett00b,knett00a,stein97,knett01b,schmi01,mulle00a,schmi03a,knett02,gruni02b,knett00e,knett01a,breni02}.
These results may serve as examples for the utility of the approach presented.

\subsection{Outlook}
We like to point out two important consequences of the formulation 
of the effective operators in second quantization.
Both implications are based on the observation that the irreducible
operators are defined on the whole Hilbert space, i.e.\ not only
for a small number of particles. The matrix elements of
the $n$-particle irreducible operators can be computed considering
only $n$-particles. But the resulting operators hold for arbitrary
number of particles.

\paragraph{Consequence 1:}
The effective Hamiltonian is valid at finite temperatures. Hence it is
possible to extend the results obtained in the first place
at {\it zero} temperature to {\it finite} temperatures.
 The technical difficulty arising  is to
treat the interactions properly, in particular the hard-core constraint.
But the description in terms of effective particles helps to
tackle this situation. Let us recall that at zero temperature no excitation,
 i.e.\ no particle, is present.
At low temperatures only a {\it small} density of particles
will be in the system. So it is well justified to use a ladder
approximation. This approximation is also suited to deal with
the hard-core constraint (Br\"uckner approach) \cite{sushk98,kotov99}. 
Note that the problems linked to the existence of anomalous Green functions
\cite{kotov99} do not occur if the particle-conserving, effective 
Hamiltonian is used. Thus the Br\"uckner approach for the {\it effective} 
Hamiltonian after a suitable mapping \cite{trebs00,knett00a}
is well justified and represents a very promising route 
to treat  finite temperatures.

\paragraph{Consequence 2:}
So far the mapping to an effective model has been constructed perturbatively.
That means that all operators, the Hamiltonian $H$, the generator $\eta$
and the observables ${\mathcal O}$, are given in a series of
some small parameter $x$. In actual applications these series are
suitably extrapolated. But a certain caveat persists if the starting
point is a local Hamiltonian. Then a calculation up to a certain order
describes processes of a certain {\it finite} range only. This restriction
can be partly overcome by extrapolating in momentum space, e.g.\
for dispersion relations $\omega(k)$. But it is difficult to
extrapolate the matrix elements of the $2$-particle irreducible interaction 
because it is not diagonal in all momenta.

This problem can be overcome by performing the continuous unitary 
transformation directly on the level of the $n$-particle irreducible
operators.  An ansatz for the effective
Hamiltonian is chosen comprising for instance all possible irreducible
$n$-particle terms and similar terms creating and annihilating particles.
This ansatz is inserted in the flow equation (\ref{flow}). Comparison of
 the coefficients $t_{j_1\ldots;i_1\ldots}$ and 
$\partial_\ell t_{j_1\ldots;i_1\ldots}$ in front of the 
terms $e^\dagger_{j_1}\ldots e^{\phantom{\dagger}}_{i_1}\ldots$
yields coupled non-linear differential equations. 
These differential equations represent renormalization equations
for the problem under study.  We call this type of transformation a 
self-similar one since the {\it kind} of terms retained stays the same.
Again, the formulation in second quantization
allows a significant generalization. We like to stress again
that the approach presented does not require a weak-coupling limit.
For illustration, however, the reader is referred to the 
 weak-coupling examples in  Refs.~\onlinecite{heidb02a,heidb02b,white02a}. 

\paragraph{Concluding Remark}
In this paper we discussed the general structure of effective operators
and a perturbative unitary transformation to derive them.
No concrete application is presented since such applications can be found
elsewhere. The two continuative points above show along which lines
the general structure can be exploited to extend the applicability
beyond zero temperature results and beyond finite range processes.
Work along these lines is partly under way, but deserves definitely
further attention.

\section{Acknowledgements}
The authors are grateful for many inspiring discussions with
M.~Gr\"uninger, H.~Monien and E.~M\"uller-Hartmann. One of us (GSU)
 acknowledges very helpful discussions with C.~Pinettes on the
problems of anomalous Green functions in the Br\"uckner approach
and its applicability at finite temperatures.

\appendix
\section{Three-Particle Irreducible Interaction}
\label{3part}
Here we complete the formulae for the irreducible 3-particle interaction
which was given in Eq.~(\ref{t_def}). The corrections $A_0$, $A_1$ and $A_2$
result from $H_0$, $H_1$ and $H_2$, respectively, as given in (\ref{rekurs}).
They read
\begin{mathletters}
\begin{eqnarray}
A_0 \!&=& E_0^A[\delta_{j_1i_1}\tilde\delta_{j_2j_3;i_2i_3}+ 
\delta_{j_1i_2}\tilde\delta_{j_2j_3;i_1i_3}+
\delta_{j_1i_3}\tilde\delta_{j_2j_3;i_1i_2}]\nonumber\\
&&\\
A_1 \!&=& t^A_{j_1;i_1}\tilde\delta_{j_2j_3;i_2i_3} +
 t^A_{j_1;i_2}\tilde\delta_{j_2j_3;i_1i_3} + 
t^A_{j_1;i_3}\tilde\delta_{j_2j_3;i_1i_2} + \nonumber\\
&&
 t^A_{j_2;i_1}\tilde\delta_{j_1j_3;i_2i_3} +
t^A_{j_2;i_2}\tilde\delta_{j_1j_3;i_1i_3}+ 
t^A_{j_2;i_3}\tilde\delta_{j_1j_3;i_1i_2}+\nonumber\\
&& t^A_{j_3;i_1}\tilde\delta_{j_1j_2;i_2i_3} + 
 t^A_{j_3;i_2}\tilde\delta_{j_1j_2;i_1i_3} +
t^A_{j_3;i_3}\tilde\delta_{j_1j_2;i_1i_2}
\\
A_2 \!&=& \delta_{j_1i_1}\tilde t^A_{j_2j_3;i_2i_3} +
 \delta_{j_1i_2}\tilde t^A_{j_2j_3;i_1i_3} + 
\delta_{j_1i_3}\tilde t^A_{j_2j_3;i_1i_2} + \nonumber\\
&&
 \delta_{j_2i_1}\tilde t^A_{j_1j_3;i_2i_3} +
\delta_{j_2i_2}\tilde t^A_{j_1j_3;i_1i_3}+ 
\delta_{j_2i_3}\tilde t^A_{j_1j_3;i_1i_2}+\nonumber\\
&& \delta_{j_3i_1}\tilde t^A_{j_1j_2;i_2i_3} + 
 \delta_{j_3i_2}\tilde t^A_{j_1j_2;i_1i_3} +
\delta_{j_3i_3}\tilde t^A_{j_1j_2;i_1i_2}
\end{eqnarray}
\end{mathletters}
where we used the shorthands
\begin{mathletters}
\begin{eqnarray}
\tilde\delta_{j_1j_2;i_1i_2} &:=&\delta_{j_1i_1}\delta_{j_2i_2}+ 
\delta_{j_1i_2}\delta_{j_2i_1}\\
\tilde t^A_{j_1j_2;i_1i_2} &:=&
 t^A_{j_1j_2;i_1i_2}+ t^A_{j_1j_2;i_2i_1} + t^A_{j_2j_1;i_1i_2}
+ t^A_{j_2j_1;i_2i_1}.\nonumber\\ &&
\end{eqnarray}
\end{mathletters}
While the actual formulae are lengthy the underlying principle
is straightforward (see main text). Note that in concrete
realizations it is often advantageous to denote only one 
representative of the states which do not change on interchange
of particles ($|ji\rangle = |ij\rangle$).
Furthermore, certain problems allow to exploit higher particular 
symmetries like spin rotation symmetry. Then additional permutation
symmetries among the various quantum numbers constituting the 
multi-index can be exploited leading to the appearance of exchange-parity
factors.


\begin{thebibliography}{10}

\bibitem{gelfa00}
M.~P. Gelfand and R.~R.~P. Singh, Adv. Phys. {\bf 49},  93  (2000).

\bibitem{uhrig98c}
G.~S. Uhrig and B. Normand, Phys. Rev. B {\bf 58},  R14705  (1998).

\bibitem{knett00b}
C. Knetter, A. B\"uhler, E. M\"uller-Hartmann, and G.~S. Uhrig, Phys. Rev.
  Lett. {\bf 85},  3958  (2000).

\bibitem{trebs00}
S. Trebst {\it et~al.}, Phys. Rev. Lett. {\bf 85},  4373  (2000).

\bibitem{heidb02a}
C.~P. Heidbrink and G.~S. Uhrig, Phys. Rev. Lett. {\bf 88},  146401  (2002).

\bibitem{heidb02b}
C.~P. Heidbrink and G.~S. Uhrig, Eur. Phys. J. B {\bf 30},  443  (2002).

\bibitem{white02a}
S.~R. White, J. Chem. Phys. {\bf 117},  7472  (2002).

\bibitem{knett00a}
C. Knetter and G.~S. Uhrig, Eur. Phys. J. B {\bf 13},  209  (2000).

\bibitem{wegne94}
F.~J. Wegner, Ann. Physik {\bf 3},  77  (1994).

\bibitem{stein97}
J. Stein, J. Stat. Phys. {\bf 88},  487  (1997).

\bibitem{mielk98}
A. Mielke, Eur. Phys. J. B {\bf 5},  605  (1998).

\bibitem{kehre97}
S.~K. Kehrein and A. Mielke, Ann. Physik {\bf 6},  90  (1997).

\bibitem{knett01b}
C. Knetter, K.~P. Schmidt, M. Gr\"uninger, and G.~S. Uhrig, Phys. Rev. Lett.
  {\bf 87},  167204  (2001).

\bibitem{schmi01}
K.~P. Schmidt, C. Knetter, and G.~S. Uhrig, Europhys. Lett. {\bf 56},  877
  (2001).

\bibitem{zheng01a}
W. Zheng {\it et~al.}, Phys. Rev. B {\bf 63},  144410  (2001).

\bibitem{fnote}
Another way to deal with
the ambiguity would be to introduce a certain ordering among the indices. Then
only one representative of the two (or more) identical states needs to be kept 
\protect\cite{knett00b,knett01b}.

\bibitem{gelfa90}
M.~P. Gelfand, R.~R.~P. Singh, and D.~A. Huse, J. Stat. Phys. {\bf 59},  1093
  (1990).

\bibitem{gelfa96}
M.~P. Gelfand, Solid State Commun. {\bf 98},  11  (1996).

\bibitem{gnote}
Depending on the details of the interaction on
the bonds it may be sufficient to consider smaller
clusters than mentioned in the main text, for
instance a pure nearest-neighbour spin exchange
reduces the range of virtual excursions. Frustration is another
mechanism which reduces the range of the effective processes, see
e.g.\ the Shastry-Sutherland model
\protect\cite{knett00b,miyah99,mulle00a}.

\bibitem{miyah99}
S. Miyahara and K. Ueda, Phys. Rev. Lett. {\bf 82}, {3701}, (1999).

\bibitem{mulle00a}
E. M\"uller-Hartmann, R.~R.~P. Singh, C. Knetter, and G.~S. Uhrig, Phys. Rev.
  Lett. {\bf 84},  1808  (2000).
  
\bibitem{zheng01b}
W. Zheng {\it et~al.}, Phys. Rev. B {\bf 63},  144411  (2001).

\bibitem{singh99a}
R.~R.~P. Singh and Z. Weihong, Phys. Rev. B {\bf 59},  9911  (1999).

\bibitem{schmi03a}
K.~P. Schmidt and G.~S. Uhrig, cond-mat/0211627 .

\bibitem{zheng03a}
W. Zheng, C.~J. Hamer, and R.~R.~P. Singh, cond-mat/0211346.

\bibitem{knett02}
C. Knetter, K.~P. Schmidt, and G.~S. Uhrig, Physica B {\bf 312-313},  527
  (2002).

\bibitem{gruni02b}
M. Gr\"uninger {\it et~al.}, J. Phys. Chem. Solids {\bf 63},  2167  (2002).

\bibitem{knett00e}
C. Knetter, E. M\"uller-Hartmann, and G.~S. Uhrig, J. Phys.: Condens. Matter
  {\bf 12},  9069  (2000).

\bibitem{knett01a}
C. Knetter and G.~S. Uhrig, Phys. Rev. B {\bf 63},  94401  (2001).

\bibitem{breni02}
W. Brenig and A. Honecker, Phys. Rev. B {\bf 65},  140407  (2002).

\bibitem{notiz1}
The coefficients $C(\underline{m})$ and $\tilde C(\underline{m})$
will be published on the web-pages 
www.thp.uni-koeln.de/\~{}gu and www.thp.uni-koeln.de/\~{}ck.

\bibitem{sushk98}
O.~P. Sushkov and V.~N. Kotov, Phys. Rev. Lett. {\bf 81},  1941  (1998).

\bibitem{kotov99}
V.~N. Kotov, O.~P. Sushkov, and R. Eder, Phys. Rev. B {\bf 59},  6266  (1999).

\end{thebibliography}

\end{document}